\def\year{2020}\relax
\documentclass[letterpaper]{article} 
\usepackage{aaai20}  
\usepackage{times}  
\usepackage{helvet} 
\usepackage{xcolor}
\usepackage{amsmath}
\usepackage{courier}  
\usepackage[hyphens]{url}  
\usepackage{graphicx} 
\def\UrlFont{\rm}  
\usepackage{graphicx}  
\usepackage[normalem]{ulem}

\title{Explaining RADAR features for detecting spoofing attacks \\ in Connected Autonomous Vehicles}
\author{
    Nidhi Rastogi\textsuperscript{\rm 1},
    Sara Rampazzi\textsuperscript{\rm 2},
    Michael Clifford\textsuperscript{\rm 3},
    Miriam Heller\textsuperscript{\rm 4},
    Matthew Bishop\textsuperscript{\rm 4}, 
    Karl Levitt\textsuperscript{\rm 4}\\
    \textsuperscript{\rm 1}Dept. of Software Engineering, Rochester Institute of Technology, Rochester, NY,\\
    \textsuperscript{\rm 2}Dept. of Computer \& Information Science \& Engineering (CISE), University of Florida, Gainesville, FL,\\
    \textsuperscript{\rm 3}Toyota InfoTech Labs, Mountain View, CA,\\
    \textsuperscript{\rm 4}Department of Computer Science, University of California at Davis, CA,\\ nxrvse@rit.edu,srampazzi@ufl.edu,michael.clifford@toyota.com,\\
    \{mheller, bishop,levitt\}@cs.ucdavis.edu
}
    
\begin{document}

\maketitle

\begin{abstract}
Connected autonomous vehicles (CAVs) are anticipated to have built-in AI systems for defending against cyberattacks. Machine learning (ML) models form the basis of many such AI systems. These models are notorious for acting like black boxes, transforming inputs into solutions with great accuracy, but no explanations support their decisions. Explanations are needed to communicate model performance, make decisions transparent, and establish trust in the models with stakeholders. Explanations can also indicate when humans must take control, for instance, when the ML model makes low confidence decisions or offers multiple or ambiguous alternatives. Explanations also provide evidence for post-incident forensic analysis. Research on explainable ML to security problems is limited, and more so concerning CAVs. This paper surfaces a critical yet under-researched sensor data \textit{uncertainty} problem for training ML attack detection models, especially in highly mobile and risk-averse platforms such as autonomous vehicles. We present a model that explains \textit{certainty} and \textit{uncertainty} in sensor input- a missing characteristic in data collection. We hypothesize that model explanation is inaccurate for a given system without explainable input data quality. We estimate \textit{uncertainty} and mass functions for features in radar sensor data and incorporate them into the training model through experimental evaluation. The mass function allows the classifier to categorize all spoofed inputs accurately with an incorrect class label.
\end{abstract}

\section{Introduction}
Machine learning (ML) applications are driving ~\cite{mcdaniel2016machine} innovation in connected and autonomous vehicles (CAVs)~\cite{barreno2010security}. To enable ”self-awareness,” CAVs are fitted with a range of perception sensors (lidar, radar, ultrasonic, camera), control sensors, communication systems (cellular, wifi, Bluetooth), control systems (velocity, steering, and braking), and ML-trained semi- and full-autonomy services. Controller area network (CAN) buses share critical sensory information with the onboard intelligent processing units ~\cite{qayyum2020securing}, or other networks, forming the backbone of self-driving car perception. These components require a continuous assessment of potential threats to safety to allow the CAVs to navigate the road. ML models used for perception, anomaly detection, and emergency maneuvers in CAVs, rely heavily upon sensor data. Research has demonstrated how these models are vulnerable to spoofing and adversarial examples resulting from small-magnitude perturbations in the input data~\cite{papernot2016limitations}. Therefore, understanding inputs (e.g., sensor data) is a critical step toward creating resilient infrastructure within which smart agents like CAVs and human actors can co-exist, thereby reducing risks to life and property~\cite{eykholt2018robust,sitawarin2018darts,he2020machine}.
\par
\textbf{Challenges}. A unique challenge from using ML-based attack detection models results from their black-box processing characteristics. ML models generate decisions that are challenging to understand or opaque, even to those of the experts who designed them. For example, experts cannot directly correlate weights assigned to sensor signals and the actual decisions. Understanding these correlations is vital to creating effective cyber threat detection and response systems since experts need to generate models in which the training (known attack and non-attack) and test (unknown attack) data may vary considerably from each other.
\par
\textbf{Motivation}. Experts need to understand when a poor decision on the part of a CAV is the result of an action by an adversary (an attack), a fault in the system (programming, design, implementation, or quality control error), or the impact of some other problem. Unfortunately, ML model opacity prevents experts from explaining these decisions swiftly. The challenge increases in a CAV since it is subject to frequent environmental changes, potential sensor performance degradation, and behavioral modification resulting from cyber attacks. Our goal is to help experts understand and improve security-related CAV ML decisions.
\par
While some decisions require \textit{certain} data, other decisions require integrating \textit{uncertain} with \textit{certain} data to justify potential actions. The integration of \textit{uncertain} data allows decision-makers to legitimize their choices in particular circumstances. When \textit{uncertain}, additional knowledge can help decision-makers increase their confidence level and decrease their margin of error, to the extent that would not be possible using only data that could be confirmed with certainty. In this paper, we define \emph{explainability} as a mechanism for reasoning about partial knowledge of \textit{uncertain} information and describing that reasoning to specific stakeholders.
\par

\textbf{Proposed Work.} Explanation combines artifacts such as models, sensors, and contextual information to describe security “events.” We present an empirical evaluation and generation of explanations that allow CAVs better perceive attacks in the face of \textit{uncertain} information. We use the Dempster-Shafer Theory (DST) of evidential reasoning \cite{shafer1976mathematical} to capture the \textit{uncertainty} in sensor data, such as delay in the validity of measured environmental data. DST is a generalization of probability theory and has previously been employed in map building for CAVs ~\cite{pagac1998evidential}, and in CAV sensor fusion ~\cite{halla2021conceptual}.

\textbf{Contributions.} In this paper, we make the following contributions:
\begin{enumerate}
    \item We propose an explanation methodology that captures the fidelity of input sensor data using the Dempster-Shafer Theory (DST) of belief functions. Using this evidence, we hypothesize that security experts can reliably train attack detection models for an autonomous vehicle, a platform enabled by a network of sensors to determine a safe and secure vehicle trajectory.
    \item Explanations describe data reliability in determining the presence of an obstacle as well as a spoofing attack by an adversary. We experimentally demonstrate a proof-of-concept implementation of the proposed explanation methodology on Radio Detection And Ranging (RADAR) sensor data. Specifically, we demonstrate through experimental evaluation that quantifying the \textit{uncertainty} in detecting an obstacle and considering this evidence to train ML-based attack models can produce more reliable attack predictions.
    \item We identify a set of challenges for providing explanations for CAV security, ranging from CAV-specific physical limitations to the vulnerability of explanations to limitations in the ability to provide meaningful explanations.

\end{enumerate}


\section{Background and Related Work}

\textbf{Sensor spoofing attacks in CAVs.} Sensor spoofing consists of manipulating a sensor's perception of the environment or its output data to generate or simulate erroneous measurements. These attacks are dangerous because they can cause unexpected changes in CAV automated driving operations, resulting in loss of steering control or sudden brake activation~\cite{6894181,nie2017free,miller2015remote}, or other problems. Sensor manipulation can also compromise detection of obstacles, road lanes, traffic lights, and signs, resulting in severe consequences for the safety of drivers, other vehicles, and pedestrians~\cite{yan2016can,cao2019adversarial,nassi2020phantom,nassi2019mobilbye}. Literature describes two main types of sensor spoofing attacks. 
\par
In the first, the attacker accesses the vehicle Controller Area Network (CAN) bus to manipulate messages sent from the sensors to the car Engine Control Units (ECUs)~\cite{wen2020plug,yang2020identify}. In the second category, the attacker injects external signals (e.g., sound, light, electromagnetic interference) to alter signals captured by the sensors to remotely manipulate vehicle behavior~\cite{yan2016can,komissarov2021spoofing,cao2019adversarial}.


\noindent\textbf{CAN bus vulnerabilities.} The CAN bus has become a common target for spoofing attacks because it is a widely used standard for in-vehicle network communications. The protocol allows the car engine control units (ECUs) to broadcast information and receive data from the CAV sensors in the form of CAN packets. Each packet contains vehicle status signals and sensor data, including vehicle velocity, acceleration, steering wheel angle, steering signal, brake~\cite{xing2019driver}. The standard CAN bus has no authentication or encryption mechanism in place. Thus, ECU cannot ensure the packets they receive are from a good source or contain altered information, which allows an adversary with access to the bus to eavesdrop on the communications and inject malicious data.

\noindent\textbf{Physical attacks on sensors.} By design, sensors are sensitive to external physical stimulus, even if the stimulus is unintended for measurement. An attacker can generate malicious physical signals transduced to altered measurements produced by the sensors. For example, attackers can inject modulated electromagnetic interference into a radar to make it perceive fake obstacles or use a magnetic field to corrupt anti-lock Braking Systems sensor measurements~\cite{yan2016can,komissarov2021spoofing,shoukry2013non,cao2019adversarial}.

\noindent\textbf{Spoofing detection.} Current spoofing data-driven detection techniques for CAVs use ML algorithms to analyze the sensor data in-vehicle communication networks. These intrusion detection systems (IDS) use large amounts of data collected from heterogeneous CAV sensors (single or multiple) and associated detection labels for training and optimization in a supervised or semi-supervised fashion. Kang et al.~\cite{kang2016intrusion} in 2016 proposed the restricted Boltzmann machine (RBM) to separate the normal from the altered CAN packets. Taylor et al.~\cite{taylor2016anomaly} developed a supervised long short-term memory (LSTM) model to predict the next packet value for a given input sequence. More recently, Zhou et al.~\cite{zhou2019anomaly} developed a ML model that learns the parameters with shared weight to improve detection. In contrast, Song et al. have proposed an IDS based on a deep convolutional neural network able to learn the vehicle network traffic patterns without hand-designed features~\cite{song2020vehicle}.
 

\noindent\textbf{Theory of Evidence.} The theory of evidence, also known as Dempster-Shafer Theory (DST), is a general framework for reasoning with \textit{uncertainty} and applies Dempster’s combination rule to combine information from different sources. DST widely applies to several tasks in the field of autonomous driving~\cite{shafer1976mathematical}. Environment perception, object tracking, classification, and decision-making tasks are examples of sources. DST creates a more accurate representation of the environment by combining evidential data. Fusing information from different sensors (e.g., car sensors data)~\cite{chavez2015multiple} is a way to combine evidence. For high-level decision-making, Clausmann et al. deploy Dempster’s rule of combination to gain a risk value for hypothesis and trajectories~\cite{claussmann2018multi}. Magnier et al. use the theory of evidence for the classification of lidar sensor data~\cite{magnier2017automotive}. As far as we know, this is the first paper on the application of DST in sensor data for explainable security.

\noindent\textbf{RADAR sensors.} Radar supports adaptive cruise control and advanced driving assistance systems (ADAS) for collision avoidance, pedestrian detection, and complement camera and lidar systems~\cite{winner2016automotive,peng2020first}. Radar sensors emit electromagnetic waves and receive the reflection to measure flight time. These sensors are frequency-modulated continuous-wave (FMCW) radars: the sensor transmits a chirp, and the time delay of the received chirp determines the distance to the reflecting object. The phase difference determines the velocity based on the Doppler effect. We consider the radar sensor as the spoofing attack target, such as in previous work~\cite{yan2016can,komissarov2021spoofing}.
\par
\noindent\textbf{Sensor uncertainty.} A CAV may need to assign different weights to different sensor inputs as sensor inputs become more or less trusted ~\cite{matei2009composite}, or produce measurements of different quality or accuracy level~\cite{thomopoulos1994sensor}. Changing environmental conditions, sensor degradation or obstruction, and spoofing attacks are potential causes of sensor uncertainty.

\noindent\textbf{Stakeholders.} The stakeholders \cite{preece2018stakeholders} for our explanations are the machine learning experts and engineers who design and manage decision-making algorithms for CAV perception, planning, and security. While passengers, vehicle operators, and third parties are potential future stakeholders in the future, researchers should address many of the challenges\ref{Challenges} listed below first.

\section{Challenges with Explanations}\label{Challenges}

Our stakeholders need explanations that address sensor \textit{uncertainty} to help them improve security and decision-making models. However, beyond this, CAVs present a unique set of challenges for explanation generation that relate to security, hard and soft timing constraints, explanation efficacy, and data availability. Note that many of these challenges generalize beyond explanations for CAVs. We elaborate these challenges below:

\begin{enumerate}
    \item \textbf{Providing Meaningful Feedback to ML and stakeholders}: How does the ML model know that it is making a wrong decision? Is there a way of providing feedback to the stakeholder regarding the ML model's decision? What kind of meaningful information do various types of stakeholders need in order to be able to react to bad ML model decisions effectively?

\item \textbf{Negative effect of erroneous explanations}: 
Incorrect, confusing, or misleading explanations may lead to poor future actions or reduce ML models' efficacy that incorporate the explanation data as input. An explanation need not be intentionally incorrect or deceptive to cause problems. It may simply be wrong but still trusted by the stakeholders that rely upon it.

\item \textbf{Attacks on Explanations}:  Explanations can be attacked by adding, deleting, or altering parts of the explanation. Similarly, attacks on model generating the explanation can lead to wrong and potentially confusing explanations. These attacks potentially put stakeholders who depend upon these explanations at risk. They can also put other connected entities at risk by stealthily incorporating attacks into ML model training datasets.

    \item \textbf{Misleading explanations}: Explanations may cause intentional misinterpretation. An explanation can mislead, deceive, hide potentially malicious activities, obfuscate, delay actions, prevent analysis, impact legal outcomes, using information that, while correct, is likely to be misinterpreted by stakeholders in a manner of an attacker's choice.
  
    \item \textbf{Missing validations for edge case explanations}: Excessive trust placed on explanations not validated earlier against edge cases. When an expert in a field provides a seemingly accurate explanation most or all of the time, others are likely to trust their explanations based on their reputations. If the expert provides an incorrect or incomprehensible explanation, their reputation can be damaged, diminishing the trust stakeholders place in the expert's future explanations. An ML model that provides correct explanations for the vast majority of scenarios is likely to earn the trust of its stakeholders. However, this does not mean that the model can explain every edge case correctly. These edge cases could result in explanations that are wrong, misleading or conceal attacks.

    \item \textbf{Real-time processing requirements}: Explanation generation in real-time for automated driving decisions has high processing requirements. CAV prediction models cannot postpone feedback concerning numerous decisions. For example, if a vehicle witnesses an obstacle, it must make decisions within milliseconds to guarantee collision avoidance. A fraction of explanations may take longer to generate than permitted time scales. Likewise, the explanation generation requirements for stakeholders may function on time scales with different orders of magnitude. Thus, generating explanations could cause some decisions to violate real-time processing requirements.

    \item \textbf{Varying temporal requirements}: The time requirements for security events in progress, or forensic analysis, may be violated by the time required to generate explanations. Explanations that take a long time to generate or analyze could violate the soft time requirements of attack detection and post-attack analysis. The longer the delay between an event and its analysis, the more the value of that analysis may diminish. Likewise, the detection of a security event may become less valuable if that detection takes place after an attack has already been completed, rather than while the attack is at a stage where a defender could stop or mitigate the attack.

   \item \textbf{Insufficient Data}: There may be insufficient data to provide a meaningful explanation. Without sufficient information, it may not be possible to generate an unambiguous explanation or a high probability of correctness.
    
\end{enumerate}

\section{Explaining Uncertainty}

We propose a novel approach for modeling attacks on connected autonomous vehicle sensors. Our approach encompasses the \textit{certainty} and reliability of various sensor signals prior to their use in attack models. Following DST of belief functions, our approach combines both detected sensor signals and the degree of belief in the sensors. The resultant belief considers the available evidence of belief from all sensors. Following DST, we allocate probability mass to sets or intervals.
\par
In this paper, we demonstrate a proof-of-concept for our approach, as applied to radar sensor spoofing attacks, using the DST classification algorithm~\cite{denoeux2008k}. DST is relevant in situations where non-random \textit{uncertainty} is present, and only a subset of sensor data contains the states that are present with certainty. It also has a distinct advantage of not requiring prior knowledge, making it particularly suitable for classifying previously unseen information. Therefore, DST can be used to classify CAV sensor signals, which are\textit{uncertain}and without precedent. Combining evidence using Dempster's Rule of Combination (DRC) achieves each sensor data item classification. In other words, DRC combines probabilities of independent items of evidence from sensor data~\cite{chen2014data}. The output label is a class with normalized probabilities according to some underlying Dempster-Shafer mass function. Each feature from sensor data computes individual mass functions. The ``hidden" mass function provides a more informative description of the classifier output than class probabilities and uses it for decision-making.

\begin{figure}[h]
 \centering
 \includegraphics[width=1\linewidth]{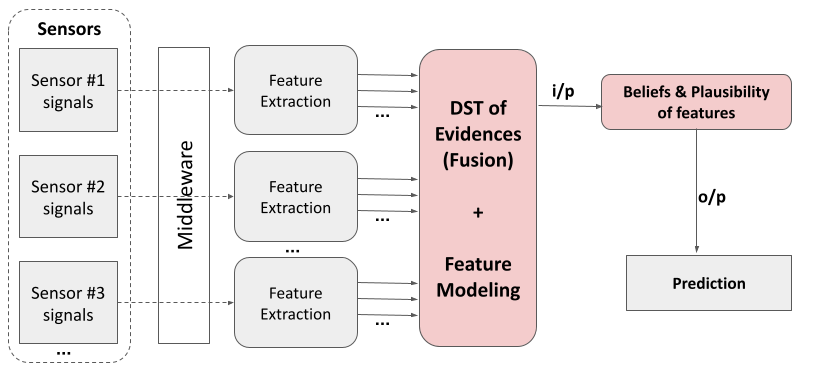}
 \caption{\textbf{The proposed explainability model}.}
 \label{fig:ExplainModel}
\end{figure}

\par
\noindent\textbf{Class Definition.} We assume the scenario of automatic obstacle tracking such as in CAV perception models~\cite{peng2020first}. We define two classes to assign the observed radar signals correlated to the perceived obstacle status - stationary \{s\}, and moving \{m\}. The set containing the two classes is called $\Theta$ and represents the \textit{frame of discernment}:
\begin{equation}
    \Theta = \{s,m\}
\end{equation}
The power set \textsf{P($\Theta$)} is the set of all possible sub-sets of $\Theta$ including the empty set $\phi$. For our problem, \textsf{P($\Theta$)} is defined as:
\begin{equation}\label{powerSet}
\textsf{P($\Theta$)} = \{\{\},\{s\},\{m\},\{s,m\}\}
\end{equation}

We postulate that the quality of evidence (signals) collected by radar is generated randomly from a Gaussian distribution with mean and variance calculated for the specific feature (see Eq.~\ref{gaussian}). We assign a ``mass" value to each element of the power set, \textsf{P($\Theta$)}. Also, ``mass" is associated with \textit{certainty} in evidence collection and is derived from radar signals.
\par
\noindent\textbf{Feature extraction and modeling.} The feature model stores the description of the characteristics mapped to two classes -- stationary \textit{s}, and moving \textit{m}. Our dataset contains features of an object in front of the radar - timestamp, density, reflection, and velocity. Description of features from the dataset is in Table~\ref{table:features}. In supervised learning, the dataset is labeled by setting the value of a flag in the radar signal. The flag uses two variables to represent the altered data in a spoofing attack or otherwise normal values for binary classification. The problem with this or a similar approach is that the malicious spoofed pattern will always match the attack detection model. In addition, the other assumption is that the data features will always be captured by the sensors in the same physical and environmental conditions as they were during model training. Especially for attack detection, this assumption can be overly stringent. Our approach is to assign instead the ``likelihood" of observing values of features such that collectively they can determine if the captured data is ``certain" (e.g., expected behavior) or not~\cite{kusenbach2020fast}.
\par
This likelihood can be modeled using normal distribution for every feature in the radar dataset as follows:
\begin{equation}\label{gaussian}
    f_{velocity} ^{\{s\}}(x) = \frac{1}{\sigma\sqrt{2\pi}}e^-\frac{(x-\mu)^2}{{2.\sigma}^2}
\end{equation}

For individual classes (stationary or moving), we have the following function, which is the summation of all features:

\begin{equation}
    f^S (x) = \sum_{A\in S} f^{A}(x)
\end{equation}

with \textit{S} $\in$ P($\Theta$). An example of \textit{S} = \{s, m\} regarding the radar signals is:

\begin{equation}
    f_{s,m} ^{\{s\}}(x) = f_{m} ^{\{a\}}(s) + f_{s} ^{\{a\}}(s)
\end{equation}
\par

\section{Classification}
In the formalism of the DST, a ``mass value" is a value between 0 and 1. It is assigned to each element of the power set P($\Theta$). DST does not regulate the method of creating these values. In our approach, these values correlate to the characteristics of the data.
\par
The mass values for the universal set, P($\Theta$), are computed for each element. Dempster's combination rule reduces the resulting mass values for each feature to one set. The lower and upper bounds of probability are computed based on the remaining set of mass values. These values, computed for each element of P($\Theta$), represent the final result. In DST's formalism, these bounds are called ``belief" and ``plausibility."

\subsection{Computing Mass Values}
To apply the Dempster-Shafer Theory, a mass \textit{m} for each element of \textsf{P}($\Theta$) is required. We use the following equation to assign the mass for a given element \textit{S} $\in$ \textsf{P}($\Theta$):

\begin{equation}\label{eq}
    m^S(x) = \frac{f^S(x)}{{\sum_{B\in P}f^B(x)}} \\
\end{equation}
with \textit{S} $\in$ P($\Theta$)

In Eq. \ref{eq}, the feature value $f^S(x)$ of one class \textit{S} is divided by the sum of the feature values of all possible classes of the power set. Therefore, we achieve a normalization where all mass values sum up to one:

\begin{equation}
    \sum_{B\in P} m^S(x) = 1 \\
\end{equation}

\subsection{Dempster's Rule of Combination}
The mass value function provides a mass value for each class and feature of the power set. For example, $m^{v}_{velocity}$ represents the mass value for the class stationary for the velocity feature. Furthermore, there are the mass values $m^{v}_{velocity}$ and $m^{r}_{reflection}$ for the velocity and reflection intensity features respectively. The following example shows how to combine several values from different features to a combined mass value $m^{v}_{combined}$. We use Dempster’s rule of combination to combine each feature's mass values sequentially.

\begin{equation}\label{eq:DST}
    m^{A}_{1,2}(x_{1,2}) = \frac{1}{1-K(x_{1,2})}\sum_{B\cap S=A\neq\emptyset}m^{S}_{1}(x_1).m^{M}_{2}(x_2) \\
\end{equation}

where,

\begin{equation}\label{eq:DST-extend}
    K(x_{1,2}) = \sum_{B\cap M=\emptyset}m^{S}_{1}(x_1).m^{M}_{2}(x_2)\\
\end{equation}

The Eq.~\ref{eq:DST} is the standard equation for Dempster’s rule of combination. It is extended to take into account the feature values $x_1$ and $x_2$ of the two features that are combined using the mass functions of these features $m_1(x_1)$ and $m_1(x_2)$. To be brief: eq. \ref{eq:DST} sums up all the mass values supporting a class. This sum is multiplied with a normalization factor $1/1-K$ (with K from eq. \ref{eq:DST-extend}). The result is a combined normalized mass value for each class.

\subsection{Calculating Belief and Plausibility}
DST uses the mass values to compute a belief (\textit{bel}) and a plausibility (\textit{pl}). The belief is the sum of all mass values supporting a class A, where each mass value is the summation of elements that are subsets of A:

\begin{equation}
    bel^A(x) = \sum_{B\subseteq A} m^B(x) \\
\end{equation}

The plausibility adds all elements related to \textit{A}. This is achieved by adding all mass values that intersect \textit{A}:

\begin{equation}
    pl^A(x) = \sum_{B\cap=A\neq\emptyset}m^{B}(x) \\
\end{equation}

\begin{figure}[h]
 \centering
 \includegraphics[width=0.9\linewidth]{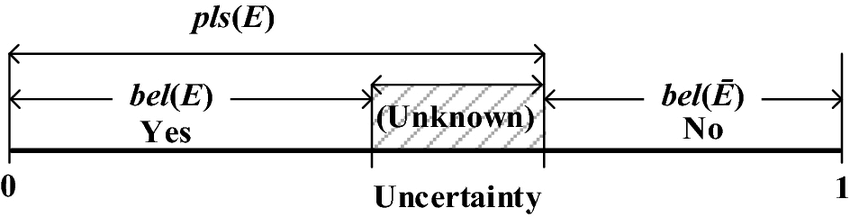}
 \caption{Belief, Plausibility, and Uncertainty \cite{klein2004sensor}}
 \label{fig:belPlau}
\end{figure}

The belief represents a lower boundary value and the plausibility of an upper boundary value for a hypothesis. Both values are in the interval between 0 and 1. The difference between them represents uncertainty. Figure~\ref{fig:belPlau} visualizes this relationship. The specialty of using the theory of evidence is that the result is comprehensive. Instead of presenting one score value for one class, our approach provides a lower and an upper bound of the belief in each hypothesis of the power set of the classes.

\begin{center}
\begin{table}
\small
\caption{RADAR signal features}
\label{table:features}
\begin{tabular}{|p{1.5cm}|p{6cm}|}
\hline
\textbf{Feature} & \textbf{Description}\\
\hline \hline
Timestamp & recorded time in seconds\\
\hline
Density & width of obstruction ahead of AV\\
\hline
Reflection & Intensity of refection from obstruction\\
\hline
Velocity & velocity of object in km/h\\
\hline
\end{tabular}
\end{table}
\end{center}

\section{Evaluation}
\par
\noindent\textbf{Attack Model Assumption.} In our scenario, we consider a sensor spoofing attack where the attacker can manipulate the radar sensor readings to change the perceived velocity of the obstacle (e.g., the victim car perceives a stationary vehicle in its trajectory when in reality the vehicle is moving far away) such as in previous work~\cite{komissarov2021spoofing}. The spoofing is achievable by external signal injection (e.g., victim vehicle behind the attacker’s vehicle with a modified radar system facing back) or tampering with the victim car to inject malicious data in the vehicle network~\cite{checkoway2011comprehensive}.
\par
\noindent\textbf{Data.} We generate a dataset that combines an existing simulated dataset\footnote{$https://github.com/lucastanger/ai\_evidence\_theory$} with our simulated dataset to match the frame of discernment. The format resembles radar sensor readings and is in human-readable format (CSV) with features described in Table \ref{table:features}. While class membership was preassigned, we used DST to evaluate class membership (stationary or moving) to each column (feature) in the dataset. Using Eq.~\ref{powerSet}, belief, plausibility and \textit{uncertainty} are created for each element of the power set, \textsf{P($\Theta$)}, see Table~\ref{table:bpufTable}. We provide individual values for each class (frame of discernment). We get 96\% accuracy when the combined feature velocity and reflection use class membership. The accuracy drops below 90\% when distance is used as the feature to assign class membership.

\begin{center}
\begin{table}
\small
\caption{Belief, Plausibility, and Uncertainty for classes -- \textit{stationary} and \textit{moving}.}
\label{table:bpufTable}
\begin{tabular}{|p{.5cm}|p{1.5cm}|p{1.5cm}|p{1.5cm}|}
\hline
 & Belief & Plausibility & Uncertainty\\
 \hline
 \hline
\textbf{s} & 0.99773 & 0 & 0.00226 \\
\hline
\textbf{m} & 0 & 0.99773 & 0.00226\\
\hline
\end{tabular}
\end{table}
\end{center}
\noindent\textbf{Detecting spoofing attacks.} We evaluate our method on a set of spoofed inputs by an attacker. We exclude detection from sensor errors or those caused by unknown factors to keep the scenario simple and focus on capturing ``uncertainty" in the signal. Belief with the highest value determines class prediction and is the accepted result corresponding to the input feature.
\par
\noindent\textbf{Analysis.} Using the theory of evidence, the number of classes increases to \textbf{$2^{(\Theta)}$} which helps us deal with scenarios where the evidence does not allow clear mapping of an input signal to one predefined class. In other words, when the CAV is under an adversarial attack, a change in feature value does not change the prediction. However, this changes the class from \textit{s} or \textit{m} to \textit{sm}. We postulate that this change in class membership is an improvement over an otherwise incorrect prediction that can change the trajectory of the CAV. Using our method, we provide more information for forensics purposes that the security experts can use to analyze the cause of the anomalous behavior.
\par
DST requires no prior knowledge of previously unseen information to detect anomalies. It can also express the value of ignorance (neither stationary nor moving). We utilize the mass assigned to different features to determine if an attacker has spoofed a data packet. Emulating an attacker in the experiment, we inject eleven spoofed packets into the dataset and flip the class membership -- for a stationary object, we assign the class as ``moving". The DST method can detect all the spoofed packets as stationary, where the mass of velocity feature is higher for moving than stationary. See results in Table \ref{table:SpoofMass}.

\begin{center}
\begin{table}
\small
\caption{Mass Functions for data points}
\label{table:massFunction}
\begin{tabular}{|p{.7cm}|p{1.5cm}|p{1.5cm}|p{1.5cm}|}
\hline
\textbf{Class} & \textbf{s} & \textbf{m} & \textbf{s, m} \\
\hline
\hline
1 & 0.347 & 0.347 & 0.306\\
\hline
2 & 0.3575 & 0.3575 & 0.2849 \\
\hline
3 & 0 & 0.625 & 0.375 \\
\hline
4 &0.312 & 0.312& 0.376\\
\hline
5 & 0 & 0.475 & 0.525\\
\hline
\end{tabular}
\end{table}
\end{center}

\begin{center}
\begin{table}
\small
\caption{Mass values for spoofed packets. All packets were correctly classified as stationary.}
\label{table:SpoofMass}
\begin{tabular}{|p{.7cm}|p{1.5cm}|p{1.5cm}|p{1.5cm}|}
\hline
\textbf{Class} & \textbf{s} & \textbf{m} & \textbf{s, m} \\
\hline
\hline
1 & 0.64 & 0 & 0.36\\
\hline
2 & 0.6976 & 0.3023 & 1.1847e-41\\
\hline
3 & 0.65& 0 & 0.35 \\ 
\hline
4 & 0.8682 & 0.1317 & 5.161e-42 \\
\hline
5 & 0.6599& 0 & 0.3400 \\ 
\hline
6 & 0.9509 & 0.0490 & 1.921e-42 \\
\hline
7 & 0.655 & 0 &0.345 \\
\hline
8 & 0.9825 & 0.0174 & 6.850e-43 \\
\hline
9 & 0.64 & 0 & 0.36 \\
\hline
10 & 0.9936 & 0.00636 & 2.494e-43 \\
\hline
11 & 0.645& 0 & 0.355 \\ 
\hline
\end{tabular}
\end{table}
\end{center}

\begin{figure}[h]
 \centering
 \includegraphics[width=1\linewidth]{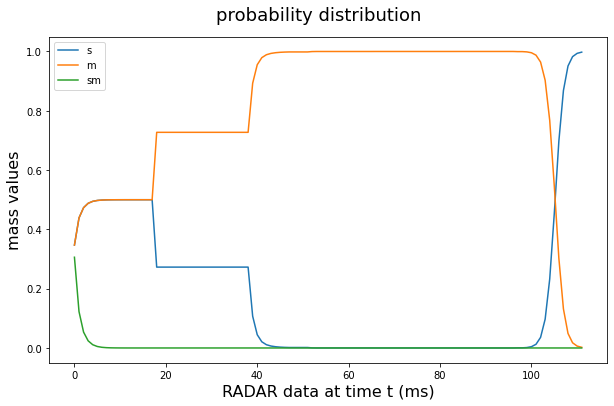}
 \caption{Probability distribution of mass functions for assigned classes (\textit{s,m,sm}) for RADAR dataset}
 \label{fig:probDistr}
\end{figure}
\vspace*{-\baselineskip}\vspace*{-\baselineskip}\vspace*{-\baselineskip}


\section{Conclusion and Future Work}
Machine Learning has become a fundamental tool for securing automated systems. However, this technology remains a black box that takes inputs and generates prediction or classification without explaining \emph{why}. Explanations are crucial for humans to decide whether an ML model decision can be trustworthy. These decisions become paramount, particularly when considering systems with high stakes that attackers can compromise. This paper investigates how explainability is essential for critical automated systems security, such as CAV. We discuss the principles, unique challenges of model explainability for security, and concrete use case described in the context of sensor spoofing attacks for autonomous vehicles. In our future work, we will develop our explainable framework to assist in complex real-time constraints such as obstacle avoidance in CAVs.

\section{Acknowledgement}
The authors would like to thank the anonymous reviewers and Praveen Chandrasekaran (RIT) for reviewing this paper and providing their invaluable feedback.

\bibliographystyle{aaai}
\bibliography{bibfile1}

\end{document}